\documentclass[structabstract]{aa}  
\usepackage[T1]{fontenc}
\usepackage{amsmath}
\usepackage{natbib}
\usepackage{graphicx}
\usepackage{txfonts}

\begin{document}
   \title{Dust production in debris discs: constraints on the smallest grains}
   \author{P. Thebault
          \inst{1}
          }
   \institute{LESIA-Observatoire de Paris, UPMC Univ. Paris 06, Univ. Paris-Diderot, France
             }
  
\offprints{P. Thebault} \mail{philippe.thebault@obspm.fr}
\date{Received ; accepted } \titlerunning{smallest grains in debris discs}
\authorrunning{Thebault}

 
  \abstract
   {The surface energy constraint puts a limit on the smallest fragment $s_{surf}$ that can be produced after a collision. Based on analytical considerations, this mechanism has been recently identified as being having the potential to prevent the production of small dust grains in debris discs and to cut off their size distribution at sizes larger than the blow-out size. }
   {We numerically investigate the importance of this effect to find out under which conditions it can leave a signature in the small-size end of a disc's particle size distribution (PSD). An important part of this work is to map out,  in a disc at steady-state, what is the most likely collisional origin for $\mu$m-sized dust grains, in terms of the sizes of their collisional progenitors.}
   {For the first time, we implement the surface energy constraint into a collisional evolution code. We consider a typical debris disc extending from 50 to 100au and two different stellar types: sun-like and A star. We also consider two levels of stirring in the disc: dynamically 'hot' (<$e$>=0.075) and 'cold' (<$e$>=0.01). In all cases, we derive $s_{surf}$ maps as a function of target and projectile sizes, $s_t$ and $s_p$,  and compare them to equivalent maps for the dust-production rate. We then compute disc-integrated profiles of the PSD and estimate the imprint of the surface energy constraint. }
  {We find that the ($s_p$,$s_t$) regions of high $s_{surf}$ values do not coincide with those of high dust production rates. As a consequence, the surface energy constraint generally has a weak effect on the system's PSD. The maximum $s_{surf}$-induced depletion of $\mu$m-sized grains is  $\sim30$\% and is obtained for a sun-like star and a dynamically 'hot' case. For the e=0.01 cases, the surface energy effect is negligible compared to the massive small grain depletion that is induced by another mechanism: the 'natural' imbalance between dust production and destruction rates in low-stirring discs identified by Thebault \& Wu(2008)}
   {}

   \maketitle
%

\section{Introduction}\label{intro}

Circumstellar debris discs have been detected around main-sequence stars because of the excess luminosity produced by $\mu$m- to mm-sized dust. Because of its relatively short lifetime, this dust cannot be primordial and is thought to be steadily replenished by a chain of erosive collisions starting from a reservoir of large, planetesimal-like, parent bodies \citep[e.g.][]{wyat08,kriv10}. In an idealized case, the size distribution of bodies within such a collisional cascade should settle toward a power law of the form $dN\propto s^{q}ds$, where the index $q$ is close to -3.5 \citep{dohn69}. Such a power law has the interesting characteristics that, while most of the disc's mass is contained in the biggest objects of the cascade, most of the geometrical cross section should be contained in the smallest grains of size $s_{min}$. This means that, at all wavelengths $\lambda \lesssim 2\pi s_{min}$, the disc's luminosity should be dominated by these smallest grains.

Determining $s_{min}$ is consequently of crucial importance. Luckily, for most stars, stellar radiation pressure imposes a 'natural' minimum cut-off size that is, in principle, easy to estimate. Indeed, because this pressure is both $\propto 1/s$ and, like stellar gravity, $\propto 1/r^2$ ($r$ being the radial distance to the star), there is a minimum size $s_{blow}$ below which radiation pressure overcomes gravity and grains are quickly  blown out from the system. Because small grains are probably produced from parent bodies on Keplerian orbits, the criteria for estimating $s_{blow}$ is given by $\beta(s_{blow}) = F_{Rad.Press.}/F_{grav.}=0.5$ (for parent bodies on circular orbits). For typical astro-silicates \citep{drai03}, values for $\beta$ range from $\sim0.5\mu$m for solar-type stars to 2-10$\mu$m for late-type A stars \citep{kriv07} \footnote{For subsolar stars, radiation pressure is too weak to overcome gravity, but stellar wind could play a similar role for active M stars like AU Mic \citep{auge06,schu15}}. As a consequence, in scattered light, disc luminosities should always be dominated by grains close to $s_{blow}$, and, for A stars, $s_{blow}$ grains should also dominate the thermal flux up to mid-IR wavelengths.

However, observations have sometimes challenged this basic view. Some discs have, for example, been found to contain important quantities of sub-micron grains below $s_{blow}$ \citep{ardi04,fitz07,john12}. While there does not seem to be one single straightforward scenario for the presence of such tiny particles, some possible explanations have been suggested. One of them is that, at very small sizes, the $\beta(s)$ curve starts to decrease with decreasing $s$ before reaching a constant value that can be $<0.5$ for K, G or F stars, thus placing sub-micron grains on bound orbits \citep{john12}. Another possibility is that we are witnessing a powerful collisional chain-reaction, called an 'avalanche', of high-$\beta$ particles that are passing through a dense ring of larger grains \citep{grig07}. 

\subsection{Discs with "too large" minimum dust sizes}\label{largesm}

Here we will focus on the opposite problem, i.e., the systems for which the observationally derived $s_{min}$ has been found to be \emph{larger} than $s_{blow}$. The existence of such systems was first unambiguously determined by \citet{pawe14}, who considered a sample of 34 resolved discs \footnote{There is an inherent degeneracy between grain size and disc radius that can only be broken for resolved systems \citep[see discussion in][]{pawe15}} and found that, while discs around A-star have $s_{min}\sim s_{blow}$, the $s_{min}/ s_{blow}$ ratio increased towards low-mass stars and could reach $\sim10$ for solar-type stars. These results were later confirmed by \citet{pawe15}, who showed that the $s_{smin}/s_{blow}$ trend is robust and does not depend on material compositions or grain porosity.

These results seem to be in contradiction with the aforementioned behaviour expected in discs where a size distribution in $dN\propto s^{q}ds$ holds down to the blow-out size $s_{blow}$, for which luminosities should be dominated by grains close to the blow-out size. Interestingly, numerical investigations of the collisional evolutions of debris discs have shown that, at least in the small-size domain, size distributions can in fact significantly depart from an idealized $dN\propto s^{q}ds$ power-law. The simulations of \citet{theb03,kriv06,theb07} have indeed shown that size-distributions can exhibit pronounced 'wavy' patterns in the $\lesssim 100s_{blow}$ domain, triggered by the absence of potential destructive projectiles below $s_{blow}$. However, this waviness cannot deprive the system from grains close to $s_{blow}$, and, generally, it even has just the opposite effect, inducing a density peak of grains in the $1.5-2s_{blow}$ region \citep{theb07}. 

A possible cause of small-grain depletion could be the well-known Poynting-Robertson drag, which causes small particles to slowly spiral starwards. However, this effect is likely to be only noticeable for tenuous discs for which the collisional timescale of small grains becomes larger than $t_{PR}$ \citep{wyat05}.
The first plausible mechanism for small-grain depletion that could also work for bright discs was proposed by \citet{theb08}, who showed that, in discs with a low dynamical excitation (low orbital eccentricities $e$ and inclinations $i$), there is an imbalance between the production and destruction rates of small grains. Indeed, while their production rate is controlled by the erosion of larger particles and is thus low because of these particles' low <$e$>, their \emph{destruction} rate, which is controlled by impacts involving the small grains themselves, is much higher because these grains are placed on high-$e$ orbits by radiation pressure regardless of the dynamical excitation in the rest of the system. As a result, depending on the value for <$e$>, there could be a strong depletion of grains up to sub-mm sizes. The main issue with this scenario is that it requires values of <$e$> that could possibly be unrealistically low (see Sec.\ref{discu}).

\subsection{Maximum surface energy and minimum fragment size}

Another, potentially more generic scenario has recently been proposed by \citet{krij14}. It is based on an energy-conservation criteria, related to the physics of collisions themselves, which, remarkably enough, had never before been invoked in the context of debris discs. The argument is that a destructive collision cannot produce a size distribution of fragments that reaches an infinitely small value, as this would require an infinite amount of energy. This is because creating new fragments means increasing the amount of exposed surface, and this requires energy.
There is thus a minimum size $s_{surf}$ in the fragment distribution, which is given by the requirement that the total surface energy of all fragments cannot exceed the kinetic energy of the impact. For the specific case of a fully destructive impact between two equal-sized bodies and for a $q=-3.5$ size distribution, \citet{krij14} derived a relation linking $s_{surf}$ to $s_{lfr}$, the size of the largest fragment, and the impactor's size $s_0$:
\begin{equation}
s_{surf}= \left( \frac{24\gamma s_{0}}{\eta\rho s_{0}v_{rel}^{2} +24\gamma} \right)^{2} s_{lfr}^{-1}
\label{smin}
\end{equation}
where $v_{rel}$ is the impact velocity, $\gamma$ is the surface energy per unit surface, and $\eta$ is the fraction of the kinetic energy that is used to create a new surface. Interestingly, this formula gives the counter-intuitive result that, for a given $v_{rel}$ and a fixed $s_{lfr}/s_0$ ratio (i.e., for self-similar impacts), the size of the smallest fragment \emph{increases} with decreasing $s_0$ sizes. 
For low-velocity impacts, low values of $\eta$ and for material with high $\gamma$ (like ice), $s_{surf}$ can be significantly higher than $s_{blow}$ for $s_0\lesssim 1\,$m \citep[see Fig.1 of][]{krij14}.

However, while deriving estimates of $s_{surf}$ is analytically possible for one given collision, giving global estimates for a whole debris disc is a much more problematic task, since the values of $s_{surf}$ will vary greatly depending on the absolute and relative sizes of the impacting objects.
In their pioneering study, \citet{krij14} attempt to derive a disc-integrated $s_{surf}$ by only considering impacts between equal-sized $s_0$ objects and by restricting themselves to "barely catastrophic" impacts where each impactor is split into two identical fragments. 
This simplifying choice relied on two main assumptions: 1) For a given target size, and if \emph{$s_{lfr}$ is fixed}, it is with a projectile of the same size that the smallest "smallest fragments" should be obtained, hence the choice of equal-sized impacts as the most constraining ones, and 2) in a collisional cascade, small grains should preferentially originate from collisions that involve particles that are barely larger than themselves; so that even if collisions amongst large objets have very small $s_{surf}$ values (because $s_{surf} \propto s_{0}^{-1}$ if $s_{lfr}/s_0$ is fixed) the amount of small dust they produce cannot compensate for the amount of small dust "not produced" (because of their larger $s_{surf}$) by collisions amongst small grains, hence the focus on the smallest "barely catastrophic" impacts . 
Under these simplifying hypotheses, the disc-integrated $s_{surf}^{DD}$ is then the $s_{surf}$ obtained for the smallest $s_0$ object that can be split in two by an impact at $v_{rel}$
\begin{equation}
\frac{s_{surf}^{DD}}{s_{blow}}= 2.4\left(\frac{r}{5\rm{AU}} \right) \left(\frac{L_{*}}{L_{\odot}} \right)^{-1} \left(\frac{f}{10^{-2}} \right)^{-2} \left(\frac{\eta}{10^{-2}} \right)^{-1} \left(\frac{\gamma}{0.1\rm{J.m^{-2}}} \right)
\label{smindd}
\end{equation}
where $f=(1.25e^2+ i^2)^{0.5}$. 

\subsection{Need for a numerical approach}

As rightfully stated by \citet{krij14} themselves, Eq.\ref{smindd} should only be taken as an order of magnitude estimate, as it only takes into account a very limited range of collision types. But even this role as an order-of-magnitude indication should be taken with caution, because the simplifying assumptions that justify the predominant role of equal-sized barely-catastrophic impacts might not hold in a realistic disc.
The first problem is that this predominant role has been inferred by assuming that $s_{lfr}$ is fixed, which is far from being the case in realistic conditions, where $s_{lfr}$ strongly depends on the two impactors' mass ratio and on $v_{rel}$ \citep[e.g.][]{lein12}.  Another problem is that it was derived assuming that a $q=-3.5$ power law holds for the whole size distribution, which several debris disc studies have shown to be erroneous, especially in the small grain-size domain. Last but not least, it neglects the contribution of \emph{cratering} impacts, which might have a dominant role in the global dust production and destruction balance \citep{theb07, koba10} \footnote{To their credit, \citet{krij14} briefly investigate the role of cratering impacts in their Appendix, but again assuming a fixed $s_{lfr}/s_{target}$ ratio (see Fig.B.1. of that paper)}.

\citet{pawe15} improved on Eq.\ref{smindd} by taking into account the fact that velocities of impacts involving small grains might have higher values because radiation pressure places small grains on highly eccentric or even parabolic orbits. This significantly reduces the values of $s_{surf}^{DD}/s_{blow}$ with respect to Eq.\ref{smindd}. Interestingly, this did not, however, improve the fit to the observationally-derived $s_{surf}^{DD}/s_{blow}$, for which the original \citet{krij14} formula seems to provide a better match \citep[see Fig.14 of][]{pawe15}. However, even the improved formula used by \citet{pawe15} is a disc-averaged analytical estimate that still relies on the same simplifying assumptions mentioned earlier.

We propose here to take these studies a step further by incorporating, for the first time, the surface-energy constraint \emph{within} a numerical collisional evolution code, estimating $s_{surf}$(i,j) for all pairs of impacting bodies of sizes $s_i$ and $s_j$ and taking these values into account in the collision-outcome prescription of our code.
We describe our numerical model and the set-ups for the different cases that we explore in Sec.\ref{model}. Results are presented in Sec.\ref{results}, where we first compare the $s_{surf}$(i,j) maps to equivalent ($s_i$,$s_j$) maps for the level of dust production in the disc. We then display the steady state particle size distributions obtained for all considered cases, and investigate the signature of the surface-energy constraint by comparing these PSDs to those obtained for control runs without the $s_{surf}$ criteria. In Sec.\ref{discu}, we then discuss the importance of the surface-energy constraint, in particular with respect to the concurrent "low-stirring" dust-depletion mechanism identified by \citet{theb08}.

\section{Model}\label{model}

We use the statistical collisional model developed by \citet{theb03} and later upgraded by \citet{theb07}. This has a "particle-in-a-box" structure, where particles are sorted into logarithmic size bins separated by a factor 2 in mass. It also has a 1D spatial resolution, being divided into radially concentric annuli. Collision rates between all size bins are computed using an estimate of the average orbital eccentricity and inclinations in each size bin. Crucially, the code takes into account the increased eccentricities, and thus impact velocities of small grains whose orbits are affected by stellar radiation pressure, as well as the fact that these grains are able to cross several concentric annuli (see Appendix of Thebault \& Augereau, 2007).
Collision outcomes are then divided into two categories, cratering and fragmentation, depending on the ratio between the specific impact kinetic energy and the specific shattering energy $Q*$, which depends on object sizes and composition. In both regimes, the size of the largest fragment and the size distributions of the other debris are derived through the detailed energy-scaling prescriptions that are presented in \citet{theb03} and  \citet{theb07}.

The main upgrade for the present runs is that we implement a prescription for $s_{surf}$. We derive $s_{surf}\rm{(i,j)}$ for all possible colliding pairs of sizes $s_i$ and $s_j$ and implement this parameter into our collision-outcome prescription: if, for a given collision between the bins "$i$" and "$j$", $s_{surf}\rm{(i,j) }> s_{min-num}$ (where $s_{min-num}$ is the smallest size bin considered in the code), we cut the post-impact fragment distribution at $s_{surf}\rm{(i,j)}$ instead of $s_{min-num}$. 

We estimate the values of $s_{surf}$(i,j) for both fragmenting and cratering impacts, using the equations presented in the Appendix of \citet{krij14}:
\begin{equation}
s_{surf}= \left[ \frac{6\gamma \left( s_{i}^{3}+s_{j}^{3}\right)^{2}} {\eta\rho v_{rel}^{2} (s_i s_j)^3} \right]^{2} s_{lfr}^{-1} \,\,\,\,\,\,\,\rm{for \,\,fragmentation}
\label{sfrag}
\end{equation}
\begin{equation}
s_{surf}= \left( \frac{6\gamma \kappa} {\eta\rho v_{rel}^{2} } \right)^{2} s_{lfr}^{-1} \,\,\,\,\,\,\,\rm{for \,\,cratering}
\label{scrat}
\end{equation}
where we assume that $i$ is the target and $j$ the projectile ($s_{i}\geq s_j$). For the cratering case, $\kappa$ is the ratio $m_{crat}/m_j$, where $m_{crat}$ is the total mass excavated from the target $i$.
The crucial point here is that, instead of having to \emph{assume} values for the largest fragment's size $s_{lfr}$ and the ratio $\kappa$, we can retrieve both quantities in a self-consistent way from our collision-outcome prescription.

\subsection{Set-up}\label{setup}

\begin{table}
\caption[]{Set-up for the numerical simulations. Here, $\gamma$ is the surface energy per surface unit of material and $\eta$ is the fraction of the kinetic energy that is used for creating a new surface.  }
\label{set-up}
\begin{tabular}{l|c}
\hline
$s_{min-num}$ & $s_{blow}$\\
$s_{max-num}$ & 50m\\
Initial mass & $0.01M_{\oplus}$\\
$r_{min}$ & 50\,AU\\
$r_{max}$ & 100\,AU\\
$L^{*}/L_{\odot}$  & 1 or 9 \\
<$e$> & 0.075 or 0.01\\
 $Q*$ prescription &  \citet{benz99} (for basalt)\\
 $s_{surf}$ & Appendix of \citet{krij14}\\
 $\gamma$ & 0.74\,J.m$^{-2}$\\
 $\eta$ & 0.01\\
\hline
\end{tabular}
\end{table}

We do not consider the full range of possible $L^{*}/L_{\odot}$ explored by \citet{pawe15}, but restrict ourselves to the two illustrative cases of a sun-like $L^{*}=L_{\odot}$ star and a $\beta$-Pic-like A5V star with $L^{*}=9L_{\odot}$.
We consider a reference debris disc extending from $r_{min}=50\,$au to $r_{max}=100$\,au. The disc's total initial mass is $M_{disc}=0.5M_{\oplus}$, distributed between $s_{max-num}=50$m and $s_{min-num}=s_{blow}$, which corresponds to $\sim 0.01 M_{\oplus}$ of $<1$mm dust\footnote{The value for $M_{disc}$ is not a crucial parameter, since it will only affect the timescale for the disc evolution without affecting the results ($s_{surf}$(i,j) and dust production maps, PSDs) obtained at steady-state}. The value of $s_{blow}$ is equal to $0.5\mu$m for the sun-like case (with 80 size bins from $s_{max}$ to $s_{blow}$) and $4\mu$m for the A star one (71 size bins).
As for the dynamical state of the disc, we consider one dynamically 'hot' case with $<e>=0.075$ and one dynamically 'cold' case with $<e>=0.01$.
For each $L^{*}$ and $<e>$ case, we run both a simulation taking into account the $s_{surf}$ prescription and a reference case with no $s_{surf}$.  
We let the runs evolve for $10^{7}$years, which is enough to reach a collisional steady-state in the dust size domain ($<1\,$cm). 

As for the free parameters of the $s_{surf}$ prescriptions (Eqs.\ref{sfrag} and \ref{scrat}), we adopt a conservative approach and chose, amongst the $\eta$ and $\gamma$ values considered by \citet{krij14}, the ones that are in principle the most favourable to yielding large $s_{surf}$ values. We thus take $\eta=0.01$ (i.e., 1\% of the kinetic energy is used for creating new surface) and $\gamma=0.74$J.m$^{-2}$, which is the value for water ice \footnote{We are aware that this value is not self-consistent with our values of $s_{blow}$, which are derived for astro-silicates. Nevertheless, this is in line with our choice of considering the maximum possible effect for the surface-energy constraint}.

All main parameters for the set-up are summarized in Table\ref{set-up}.

\begin{table}
\caption[]{Main results for all four collisional runs. $s_{surf}^{DD}/s_{blow}$ is the simplified analytical value given by Eq.\ref{smindd}. $\Delta M_{s\leq2s_{blow}}$ and $\Delta M_{4s_{blow}\leq s\leq20s_{blow}}$ are the fractional excess (or depletion) of 'small-' and 'medium'-sized dust, respectively, as compared to a control simulation with no constraint on $s_{surf}$.}
\label{resu}
\begin{tabular}{c|c|c|c|c|c|}
\hline
$L^{*}/L_{\odot}$ & <$e$> &  $s_{blow}$ &  $s_{surf}^{DD}/s_{blow}$ & $\Delta M_{s\leq2s_{blow}}$ & $\Delta M_{4s_{blow}\leq s\leq20s_{blow}}$\\
\hline
9 & 0.075 & 4$\mu$m & 0.4 &  -0.038& +0.074\\
9 & 0.01 & 4$\mu$m & 19.8 & -0.049 & +0.199\\
1 & 0.075 & 0.5$\mu$m &3.1 &-0.277 & +0.410\\
1 & 0.01 & 0.5$\mu$m &153 &-0.127& +0.145\\
\hline
\end{tabular}
\end{table}

\section{Results}\label{results}

\subsection{Dust progenitors and $s_{surf}$ maps}\label{maps}

\begin{figure*}
\makebox[\textwidth]{
\includegraphics[scale=0.5]{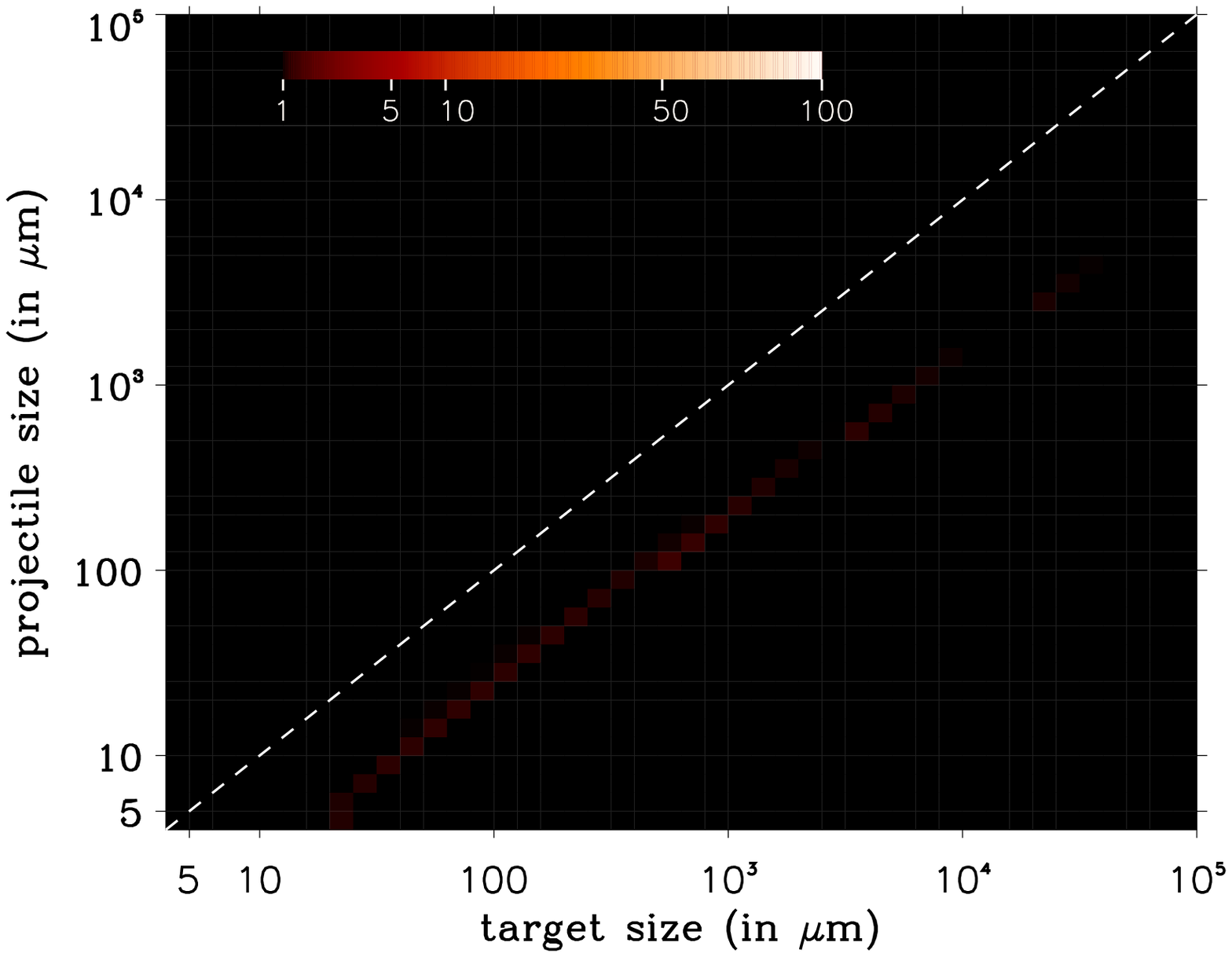}
\includegraphics[scale=0.5]{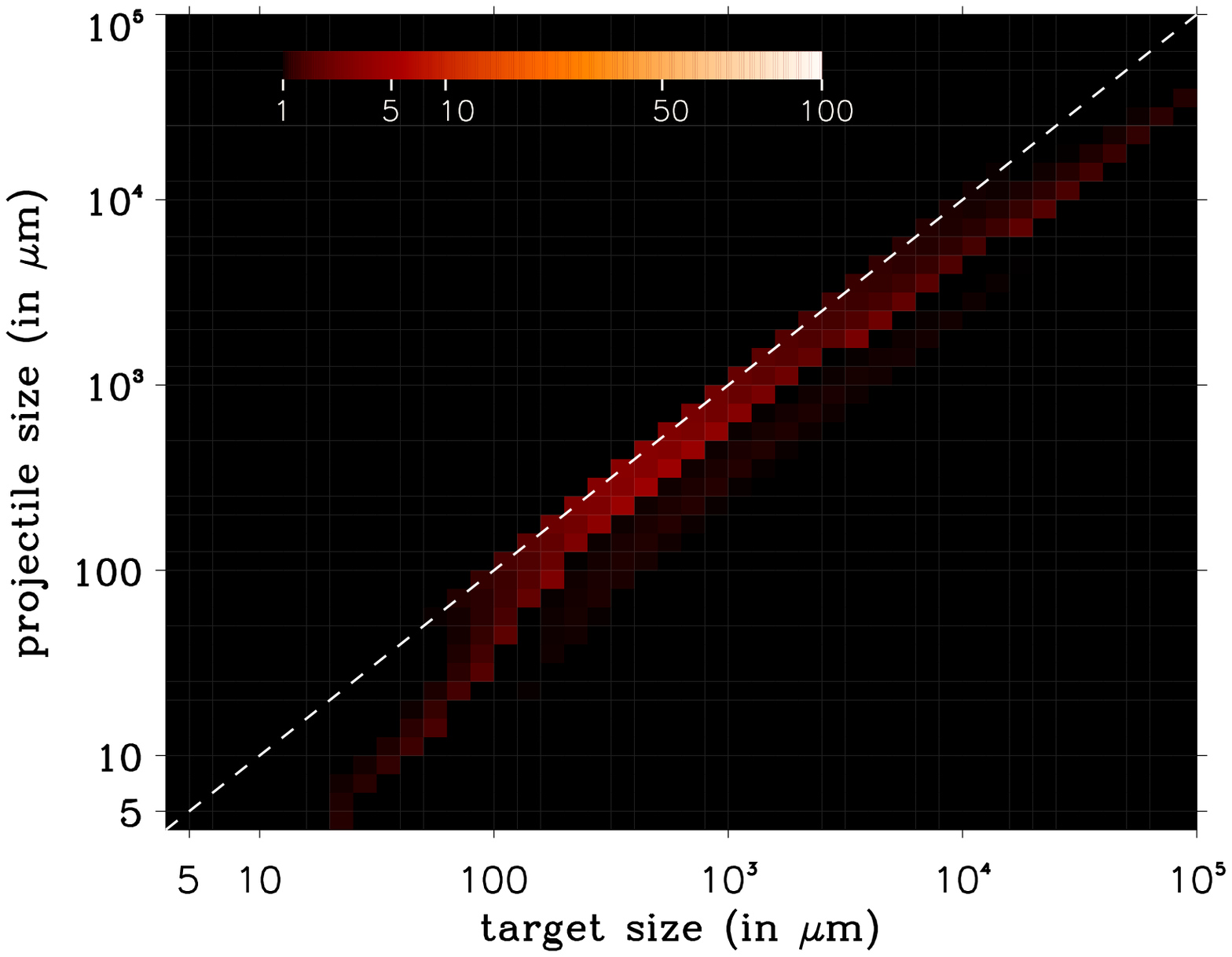}
}
\makebox[\textwidth]{
\includegraphics[scale=0.5]{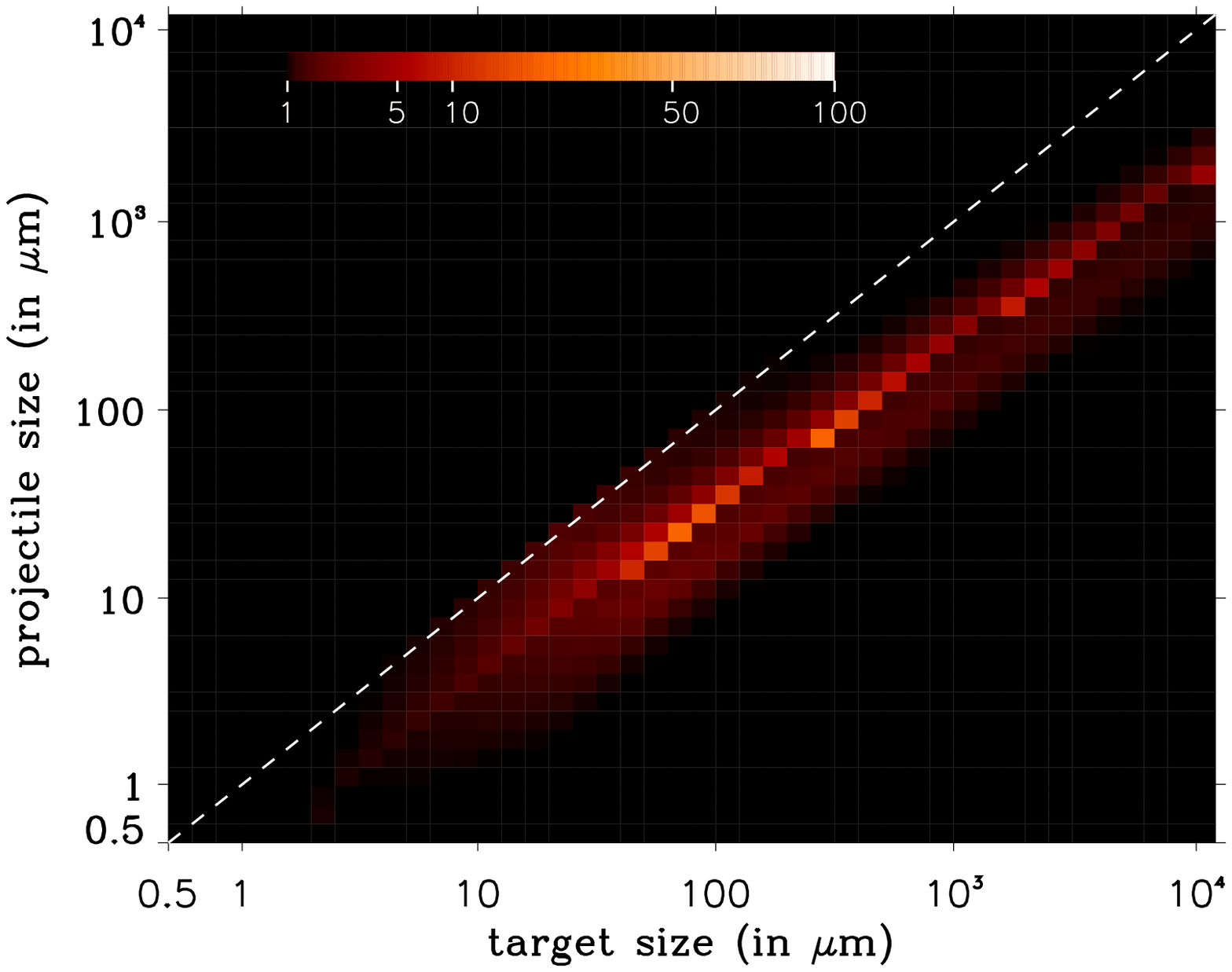}
\includegraphics[scale=0.5]{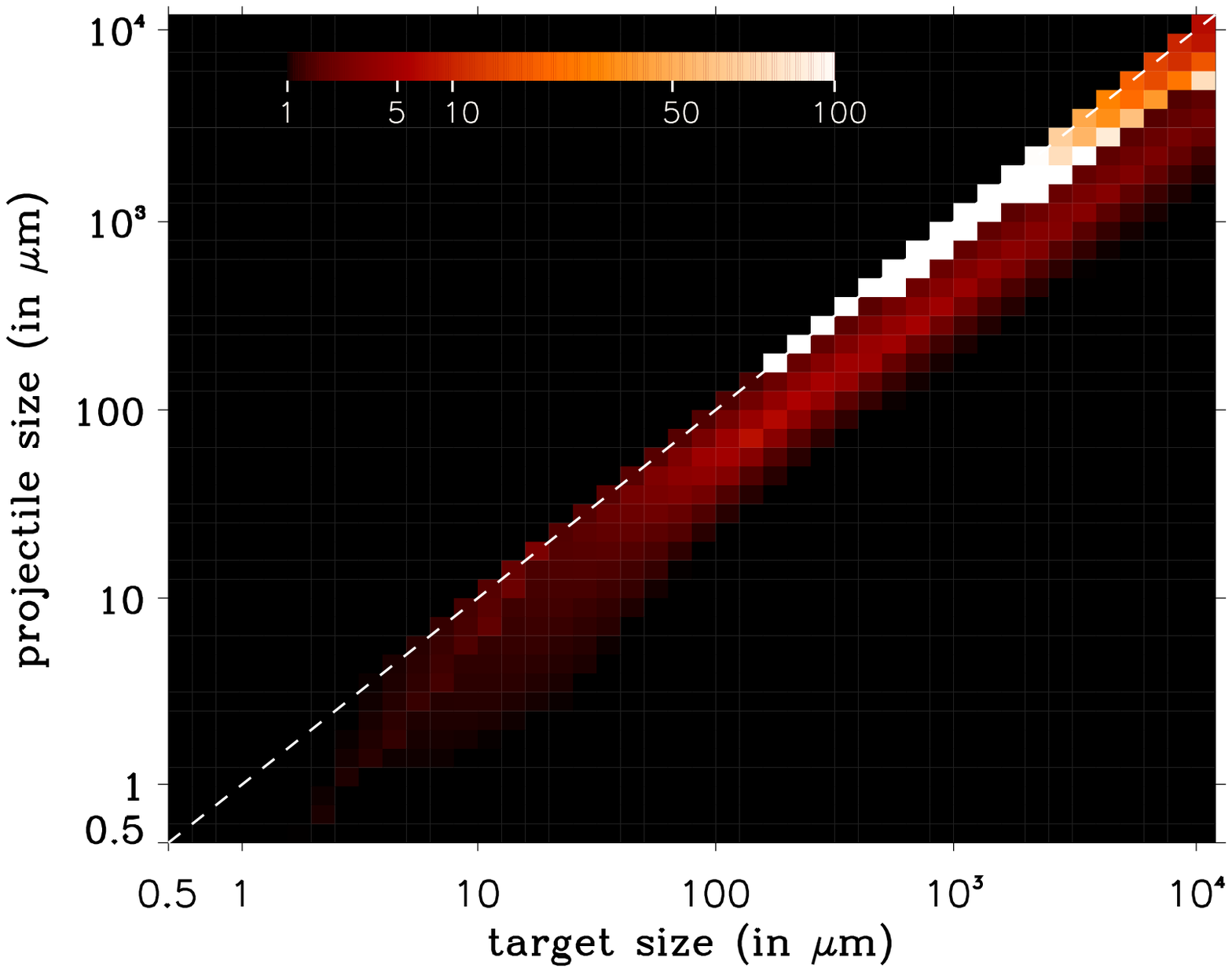}
}
\caption[]{Disc-integrated mean values of the $s_{surf}/s_{blow}$ ratio for all possible target-projectile pairs of sizes $s_i$ and $s_j$. \emph{Upper left panel}: A-star and <$e$>=0.075. \emph{Upper right}: A-star and <$e$>=0.01. \emph{Bottom left}: solar-type star and <$e$>=0.075. \emph{Bottom right}: solar-type star and <$e$>=0.01. The dashed white line delineates the $s_i=s_j$ diagonal of equal-sized impactors.}
\label{sminmap}
\end{figure*}

\begin{figure*}
\makebox[\textwidth]{
\includegraphics[scale=0.5]{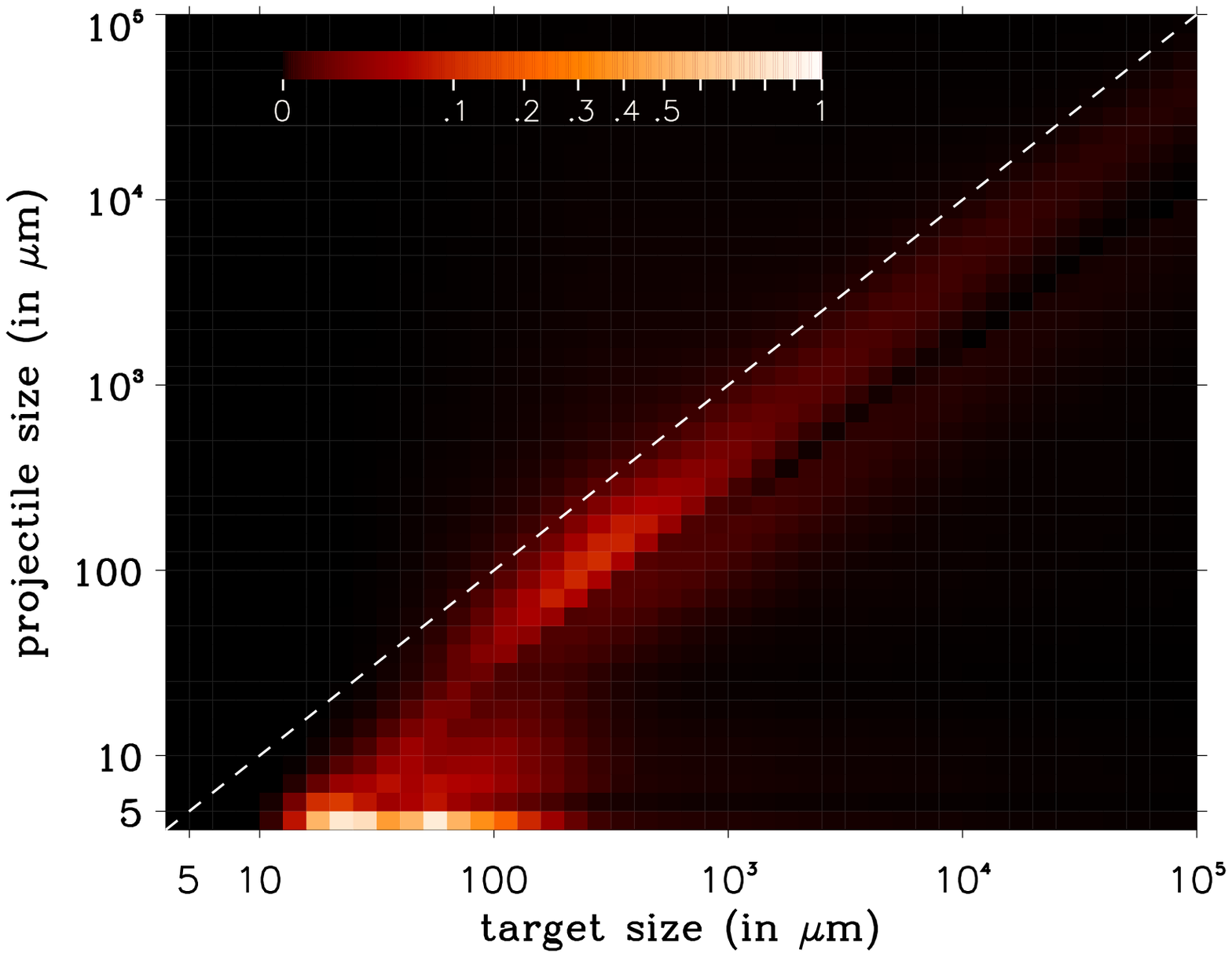}
\includegraphics[scale=0.5]{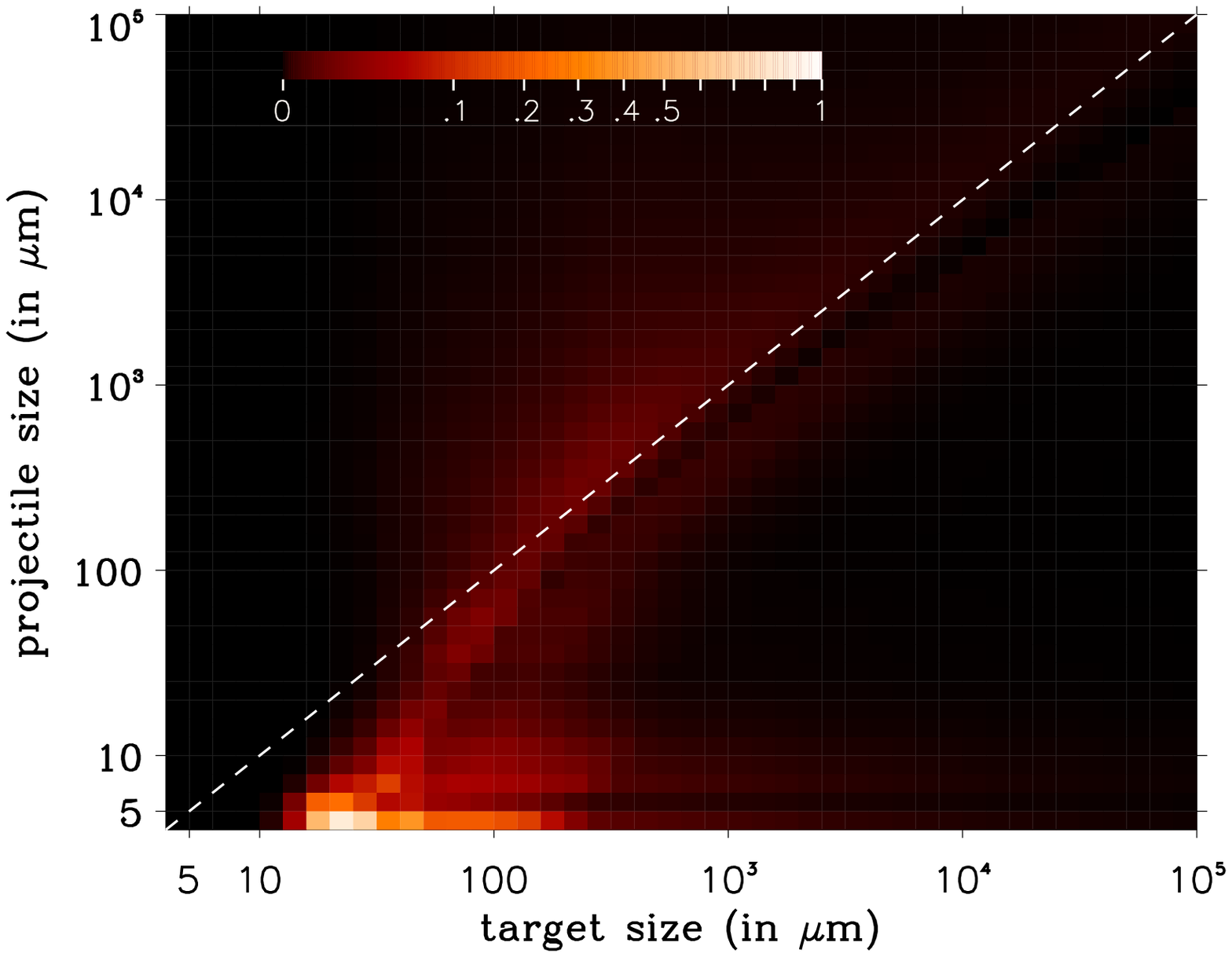}
}
\makebox[\textwidth]{
\includegraphics[scale=0.5]{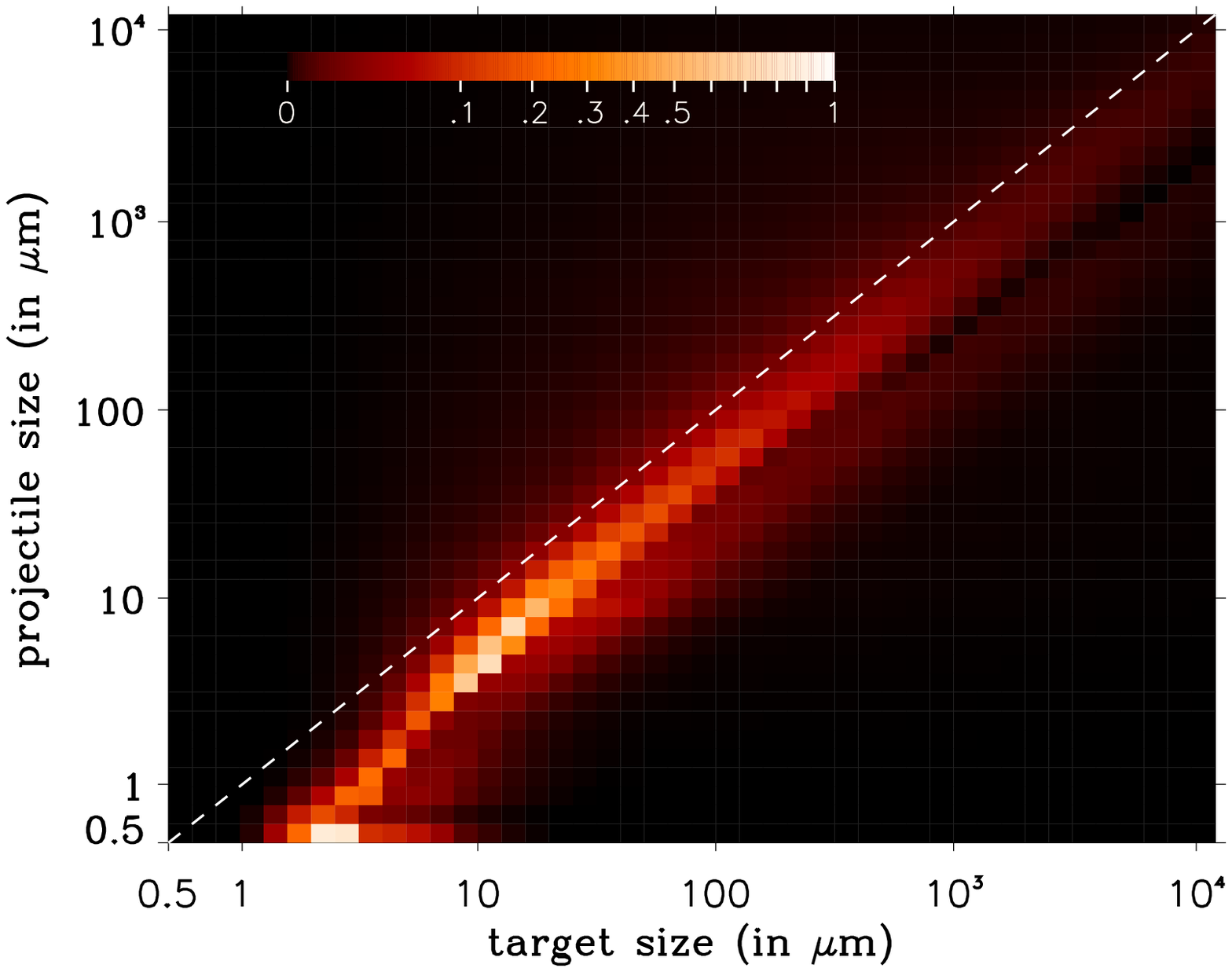}
\includegraphics[scale=0.5]{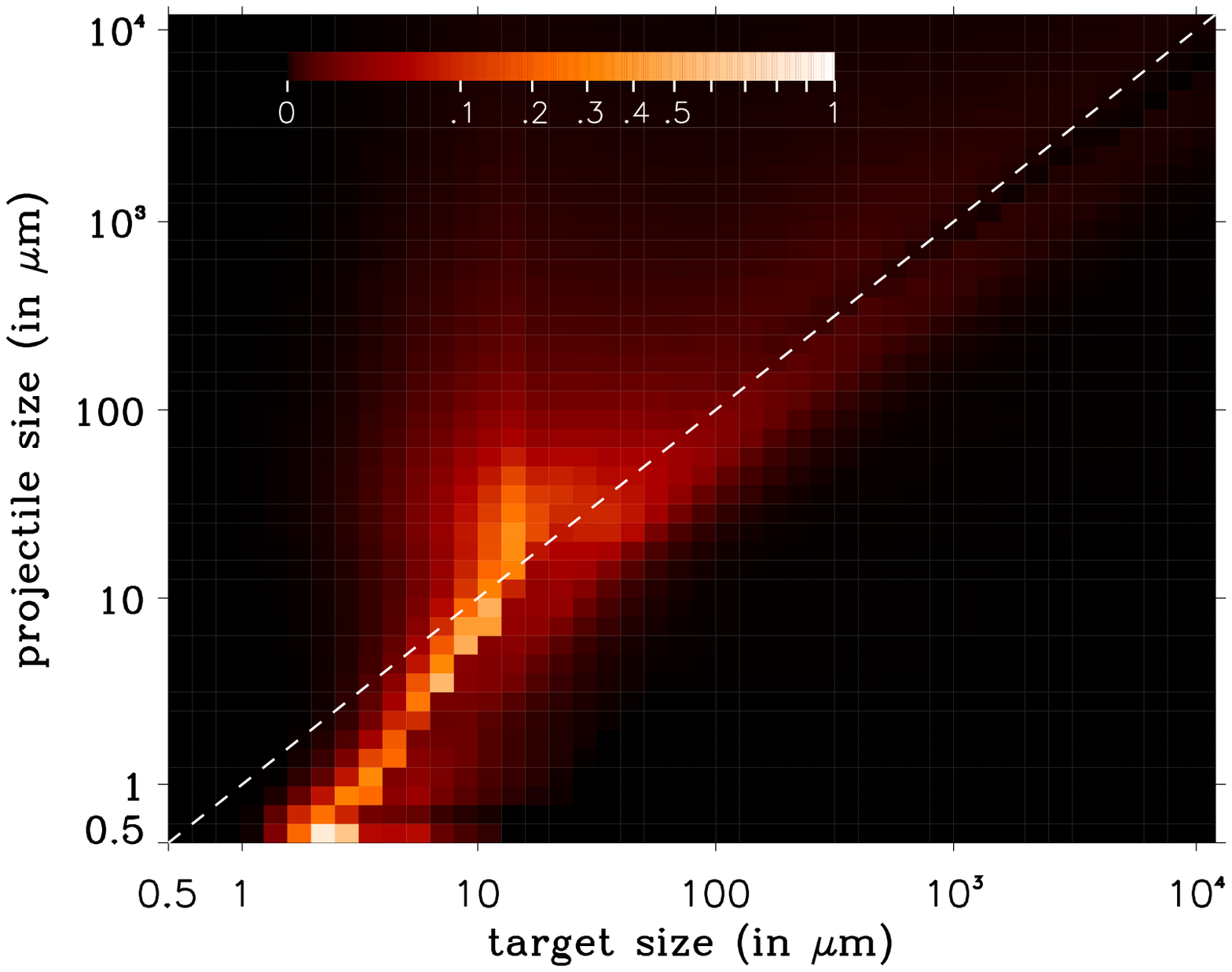}
}
\caption[]{Disc-integrated and normalized production rates of $s\leq10\mu$m dust as a function of target and projectile sizes. These maps are estimated, at collisional steady-state, for the 'control runs' where the $s_{surf}$ criteria is turned off. This allows for an easy comparison with Fig.\ref{sminmap} by directly showing which high $s_{surf}$(i,j) regions will have a significant damping effect on the small-dust production rate.
\emph{Upper left panel}: A-star primary and <$e$>=0.075. \emph{Upper right}: A-star primary and <$e$>=0.01. \emph{Bottom left}: solar-type primary and <$e$>=0.075. \emph{Bottom right}: solar-type primary and <$e$>=0.01. }
\label{prodmap}
\end{figure*}

For each of the considered set-ups, Fig.\ref{sminmap} presents the $s_{surf}\rm{(i,j)}$ value for every pair of impacting target and projectile of sizes $s_i$ and $s_j$. We also show, as a useful comparison tool, the equivalent ($s_i$,$s_j$) maps of the amount of  $\mu$m-sized dust produced as a function of target and projectile sizes (Fig.\ref{prodmap}). 

\subsubsection{A-star}

For the dynamically 'hot' (<$e$>$=0.075$) A-star case, Eq.\ref{smindd} predicts a low $s_{surf}^{DD}/s_{blow}$ of only $\sim 0.4$, and our $s_{surf}\rm{(i,j)}$ map is in good agreement with this prediction, since almost all (i,j) impacts result in $s_{surf}/s_{blow}<1$ (Fig.\ref{sminmap}a) . There is, admittedly, a very narrow range of impacts for which $s_{surf}/s_{blow}\geq 1$, but even there the ratio does not exceed 2. Moreover, this limited family of $s_{surf}>s_{blow}$ collisions corresponds to a region that has only a very limited contribution to the total dust production. The dust production is instead dominated by impacts that involve $s\sim s_{blow}=5\mu$m projectiles and larger $5s_{blow}\lesssim s \lesssim 20s_{blow}$ targets (see Fig.\ref{prodmap}a). This is because $s\sim s_{blow}$ grains are placed on high-$e$ orbits by radiation pressure and impact all other grains at very high velocities. They are thus very efficient at fragmenting targets and producing dusty debris over a wide range of target sizes.

The situation is slightly different for the dynamically 'cold' A-star case, for which the fraction of $s_{surf}$>$s_{blow}$ impacts is more extended and the maximum value for  $s_{surf}\rm{(i,j)}$ reaches $\sim 5$ (Fig.\ref{sminmap}b). This value is, however, well below the analytical prediction $s_{surf}^{DD}\sim 20s_{blow}$ of Eq.\ref{smindd}. We note that these high-$s_{surf}$(i,j) impacts are located close the $s_i=s_j$ diagonal, which could appear to contradict Fig.B.1 of \citet{krij14}, which shows instead a \emph{minimum} of the $s_{surf}$ curve for equal-sized impactors. But this is because this analytically-derived curve is only valid for \emph{fragmenting} impacts, whereas here, with <$e$>$=0.01$, the high-collisional-energy requirement for fragmentation can only be met when projectiles have sizes comparable to the target. All other cases result in cratering, for which $s_{surf}$ are, in general, much smaller. As was the case for the <$e$>$=0.075$ run, the (i,j) region of high $s_{surf}$ values does not match the region of highest dust production, which corresponds, here again, to fragmenting impacts by $s\sim s_{blow}$ projectiles on $5s_{blow}\lesssim s \lesssim 20s_{blow}$ targets (Fig.\ref{prodmap}b). Nevertheless, the contribution of the high- $s_{surf}$ region is not fully negligible, so that we expect the maximum surface energy criteria to have at least some effect on the system's evolution.

\subsubsection{Solar-type star}\label{sunsmin}

For a Sun-like star, orbital velocities are lower and collisions less energetic than for an A-star for equivalent orbital parameters. The $s_{surf}/s_{blow}$ values are thus logically higher, as is clearly seen in Fig.\ref{sminmap}c and d. The $s_{surf}/s_{blow}>1$ domain is much more extended than for an A star, with peak values reaching $\sim 25$ for the <$e$>=0.075 case and even $\geq 100$ for the dynamically cold case. As was already the case for the A-star runs, these peak $s_{surf}$ values are located, in the ($s_i$,$s_j$) map, on a line corresponding to the transition from cratering to fragmenting impacts \footnote{This relatively sharp transition at the cratering/fragmentation boundary is due to the discontinuity in the $s_{surf}$ prescriptions of \citet{krij14} (Eqs.3 and 4) at this boundary ($s_{lfr}=2^{-1/3}s_1$). More refined and self-consistent $s_{surf}$ prescriptions should probably be derived in the future, but we chose to stick to the laws given by \citet{krij14} for this exploratory work.} . For the <$e$>=0.075 run, this transition always occurs for $s_j \leq 0.5s_i$, with a $s_j/s_i$ ratio that is decreasing towards larger sizes following an approximate law in $s_2/s_1 \sim 0.4 - 0.1$log$(s_2/10\mu$m$)$\footnote{This negative slope reflects the fact that, in the strength regime, the specific shattering energy $Q*$ decreases with increasing sizes \citep{benz99}}. For <$e$>=0.01, the peak $s_{surf}$ line still corresponds to the fragmentation/cratering transition, but is now much closer to the $s_i=s_j$ diagonal, which reflects the fact that, for these lower impact velocities, it takes a larger projectile to fragment a given target. 

We also note that, for both the <$e$>=0.075 and <$e$>=0.01 cases, there are no high $s_{surf}/s_{blow}$ values for impactors $\lesssim 10\mu$m. This is because such small grains have more energetic impacts (able to produce smaller fragments) because of their radiation-pressure-affected orbits. Incidentally, this $s\lesssim 10\mu$m region is the one where the $\mu$m-dust production is the highest (Figs.\ref{prodmap}c and d). This means that a large fraction of the dust-generating impacts are, in fact, able to produce fragments down to the blow-out size. The regions of  high $s_{surf}/s_{blow}$>5 only contribute to approximately 20\% of the $\leq10\mu$m dust production for both the <$e$>=0.075 and <$e$>=0.01 cases.

\subsection{Size distribution}

\begin{figure}
\includegraphics[scale=0.5]{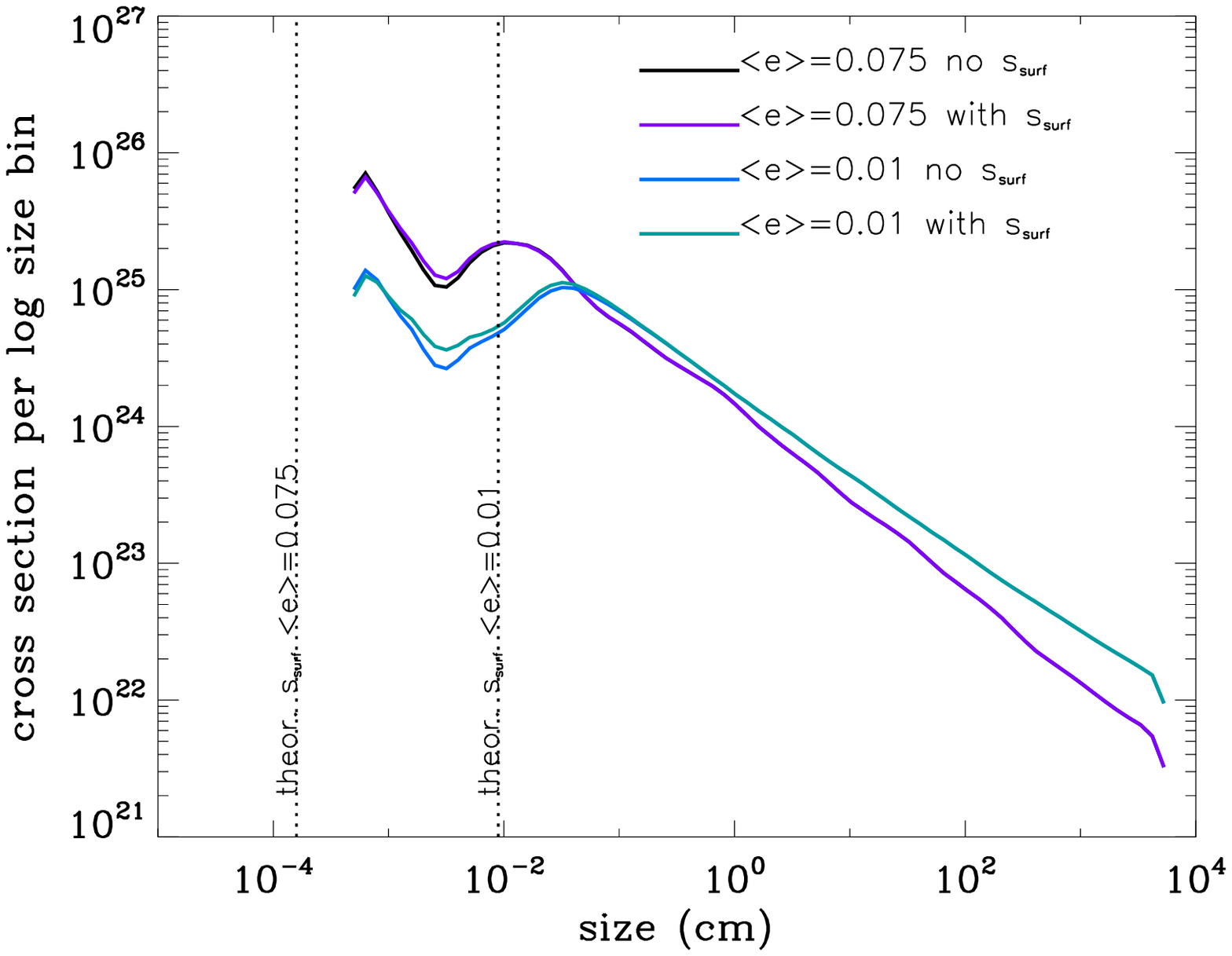}
\caption{A-star case. Geometrical cross-section as a function of particle size, at $t=10^{7}$\,years, integrated over the whole 50-100\,au disc, for both the dynamically 'hot' (<$e$>=0.075) and 'cold' (<$e$>=0.01) cases. For each case, a reference run with the $s_{surf}$ constraint switched-off is also presented. The two vertical dotted lines represent the analytical values given by Eq.\ref{smindd}.}
\label{sigastar}
\end{figure}

\begin{figure}
\includegraphics[scale=0.5]{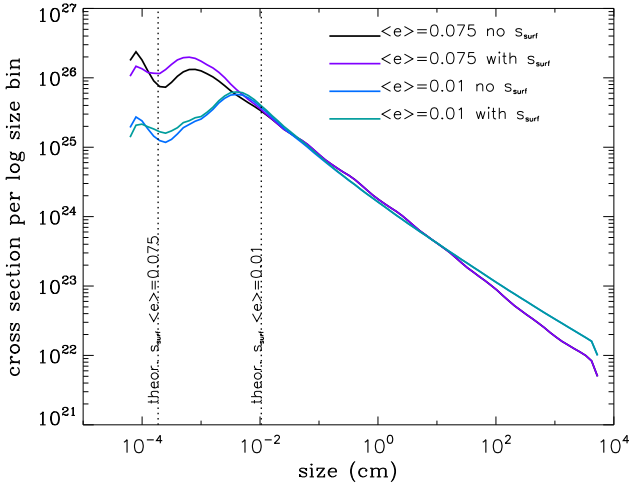}
\caption{Same as for Fig.\ref{sigastar}, but for a solar-type star.}
\label{sigsun}
\end{figure}

Figures \ref{sigastar} and \ref{sigsun} present the particle size-distribution (PSD), in terms of the differential distribution of geometric cross section $d\Sigma/ds$, at collisional steady-state\footnote{Or, more exactly, when steady-state is at least reached where it matters here, that is in the $s$<1cm domain. In the large-size domain, there is a difference of PSD slopes between the dynamically hot and cold discs that is due to the  <$e$>=0.01 systems not yet having had time to reach collisional steady-state for these larger objects. Collisional steady-state does indeed work its way up the PSD with time, at a rate that strongly increases with decreasing dynamical stirring \citep[e.g.,][]{lohn08}.}, for all four of the explored set-ups, as well as for the control runs where the $s_{surf}$ constraint is ignored.
Visualizing the $d\Sigma/ds$ distribution allows us to immediately identify which particle sizes dominate the system's optical depth, and thus its luminosity.

\subsubsection{Generic features}\label{genres}

Before assessing the additional effect of the $s_{surf}$ constraint, let us first emphasise some characteristics of the obtained size distributions that are well known generic features of steady-state collisional discs, and that also affect the small-size domain of the PSDs. 

A first 'classic' result is that, for all cases, the size distribution displays a clear wavy structure in the $\lesssim 100\mu$m domain. This waviness is more pronounced than in \citet{pawe15} because we consider a wider disc, so that small grains, placed on high-$e$ orbits par radiation pressure, will be able to impact target particles at a higher impact angle and thus higher $v_{rel}$. This will reinforce their shattering power and thus the amplitude of the size distribution wave \citep[see][for a detailed discussion on "wave-enhancing" factors]{theb07}. 

Another crucial result is the strong depletion of small particles for all <$e$>=0.01 cases. This is the signature of the mechanism identified by \citet{theb08}, i.e., the strong imbalance between the small-dust production and destruction rates for dynamically cold systems (see Sect.\ref{largesm}). As underlined by \citet{pawe15}, this effect is more pronounced for solar-type stars, for which we find a clear drop of the $d\Sigma/ds$ curve at $s\lesssim50s_{blow}$, than for A stars, where the PSD is roughly flat in the same $s\lesssim50s_{blow}$ domain. Nevertheless, for both cases the depletion of small grains is very strong, exceeding one order of magnitude as compared to a dynamically 'hot' system.

\subsubsection{Effect of the $s_{surf}$constraint}

As expected from the $s_{surf}\rm{(i,j)}$ and dust-production maps (see Sec.\ref{maps}), the effect of the $s_{surf}$ constraint is very weak for the A-star runs. For the <$e$>=0.075 case, the PSD is almost indistinguishable from the control run with $s_{surf}$ switched off (Fig.\ref{sigastar}). Some differences between the with- and without-$s_{surf}$ runs are visible for the <$e$>=0.01 case, but they remain relatively limited. Interestingly, they are not so much visible as a dearth of grains in the $s\leq10-20\mu$m domain where the few $s_{surf}\geq s_{blow}$ values lie (see Fig.\ref{sminmap}b), but rather as an \emph{excess} of larger particles in the $\sim$30-50$\mu$m range. This is a direct consequence of the specific dynamics, and thus destructive efficiency, of grains close to $s_{blow}$.
In the "$s_{surf}$-free" case, it is indeed this high destructive efficiency that is responsible for the strong depletion of $30-50\mu$m particles in the PSD (the first dip in the "wave")\footnote{Note that the presence of a wave does not depend on the high-$e$ orbits of small grains, but these high-$e$ orbits, and thus high impact velocities, strongly increase the wave's amplitude}. So removing a small number of small destructive impactors will necessarily increase the number of $30-50\mu$m grains. And given that we start from a strongly depleted population of $30-50\mu$m particles in the $s_{surf}$-free case, their \emph{relative} increase in the $s_{surf}$ runs will be higher than the relative decrease of $s\leq10-20\mu$m grains.
And this is exactly what we observe (see Table \ref{resu}): a $\sim 5$\% depletion of $s\sim s_{blow}$ grains that causes a $\sim20$\% excess of $s\sim30\mu$m particles, which is the size of the biggest objects that can be destroyed by projectiles close to $s_{blow}$. However, for this <$e$>=0.01 case, the $s_{surf}$-induced depletions and excesses remain marginal when compared to the much more significant global depletion of $s\lesssim50s_{blow}$ grains caused by the low dynamical excitation of the system (see Discussion).

The imprint of the $s_{surf}$ limit is, logically, much more visible around a Sun-like star (Fig.\ref{sigsun}). Interestingly, this is especially true of the dynamically hot disc, for which the depletion of $s\leq 2s_{blow}$ grains is close to 30\%, and the related excess of $4s_{blow}\leq s\leq20s_{blow}$ particles exceeds 40\% (Table \ref{resu}). For the <$e$>=0.01 run, these excesses and depletions are only $\sim12$\% and $\sim14$\%, respectively. This is a rather counter-intuitive result, as Eq.\ref{smindd} predicts a $s_{surf}^{DD}/s_{blow}$ ratio that is 50 times higher for <$e$>=0.01 than for <$e$>=0.075, and even the more accurate Figs.\ref{sminmap}c and d show peak $s_{surf}$(i,j) values that are still 6-7 times higher for the dynamically cold case. However, as discussed in Sec.\ref{sunsmin}, the crucial point is that these ($s_i$,$s_j$) regions of high $s_{surf}$ values do $not$ match those of high dust production. Most of the dust is indeed created by collisions on $s\leq10\mu$m ($\lesssim20s_{blow}$) targets (Fig.\ref{prodmap}), for which $s_{surf}$ rarely exceeds $s_{blow}$ (Fig.\ref{sminmap}). 
Another reason for which the $s_{surf}$ constraint only leaves a weak signature on the <$e$>=0.01 system is that, in the $s\leq100\mu$m range, the PSD is already massively depleted because of the dust production imbalance inherent to dynamically 'cold' discs. This effect is even stronger than for the A-star case, with a depletion that reaches almost two orders of magnitude for $s\sim s_{blow}$ grains (Fig.\ref{sigsun}).

\section{Discussion and conclusion}\label{discu}

Figs.\ref{sigastar} and \ref{sigsun} seems to indicate that the two dynamically 'hot' cases bear some similarities with the prediction of Eq.\ref{smindd} regarding the imprint of the $s_{surf}$ constraint on the PSD. In the A-star case, this similarity is, of course, simply that the surface energy constraint has no effect on the PSD, but in the <$e$>=0.01 case, we do indeed find a depletion of grains in the $s\leq 2s_{blow}$ domain, which is roughly consistent with the $s_{surf}^{DD}/s_{blow}\sim 3$ value given by Eq.\ref{smindd}. 
However, this quantitative agreement is probably largely a coincidence, firstly because the depletion of $s\leq s_{surf}$ grains does not come from the "smallest-barely-catastrophic-equal-size-impactors" collisions considered in deriving Eq.\ref{smindd} (see Figs\ref{sminmap}c and \ref{prodmap}c), and secondly because this depletion stops at $s\sim 2s_{blow}$ largely because of the 'natural' waviness of the PSD, combined with the \emph{excess} of $\geq 3s_{blow}$ particles.
We also note that this depletion is limited to $\sim 30$\% and does not create a sharp cut-off in the size distribution. This limited amplitude, combined to the aforementioned excess of particles in the $\sim 3s_{blow}$ to $\sim$20-30$s_{blow}$ range, renders the PSD plateau-like at sizes smaller than $\sim10s_{blow}$. It is thus difficult to define a proper $s_{min}$ for the size distribution, but, at least from a qualitative point of view, we confirm that, for this case, the maximum energy criteria does have a visible effect on the lower-end of the PSD.

The situation is radically different for the dynamically cold cases, for which the simulated size distributions are much less affected by the $s_{surf}$ constraint than would be expected from simple analytical estimates, which predict $s_{surf}^{DD}/s_{blow}$ values in excess of 20, or even 150 (see Table \ref{resu}). Even though we do find some ($s_i$,$s_j$) target-projectile configurations that result in large $s_{surf}$ comparable to these values (Fig.\ref{sminmap}), the decisive point is that these collisions only have a marginal contribution to the disc's total small-dust production.
As a result, the differences with the reference  $s_{surf}$-free cases' PSDs remain very limited: for neither the A-star nor the Sun-like cases do we obtain a depletion that exceeds $\sim10$\% in the small grain domain close to $s_{blow}$. Crucially, the $s_{surf}$ constraint is never strong enough to change the size of grains that dominate the system's geometrical cross-section, and thus its luminosity. 

More importantly, we confirm that the most efficient way of depleting a collisional debris disc from its small grains is, by far, to reduce its dynamical excitation. The production/destruction imbalance mechanism identified by \citet{theb08} for low-stirring discs has an effect that exceeds, by more than one order of magnitude, that of the $s_{surf}$ limit. This is true both for the amplitude of the small-grain depletion and for the size-range that is affected. For low <$e$> values, the size $s_{cold}$ below which the PSD is depleted is, to a first order, given by the relation
\begin{equation}
\frac{\beta(s)}{1-\beta(s)} = <e>
\label{thebwu1}
\end{equation}
which translates into
\begin{equation}
s_{cold} = \frac{1+\rm{<e>}}{2\rm{<e>}} \,\,\,s_{blow}
\label{thebwu2}
\end{equation}
leading to $s_{cold}\sim 50s_{blow}$ for our <$e$>=0.01 case, a value that roughly agrees with the one obtained in the simulations (Figs.\ref{sigastar} and \ref{sigsun}).
It should be noted that this dominance of the low-stirring imbalance effect over the surface energy constraint is probably even stronger than that witnessed in Figs \ref{sigastar} and \ref{sigsun}, since we have taken, for our $s_{surf}$ prescription, the parameters that were the most likely to lead to high $s_{surf}$ values (see Sec.\ref{setup}).
On a related note, choosing disc configurations that enhance the amplitude of the surface-energy constraint, by, for instance, decreasing <$e$> or increasing [$r_{min},r_{max}$], would not change this dominance either, because these configurations would \emph{also} enhance the dust-production-imbalance effect by further decreasing the level of stirring in the disc. 

We emphasise that we did not attempt to fit observed $s_{min}/s_{blow}$ trends with an exhaustive parameter exploration (<$e$>, $M_{disc}$, $[r_{min},r_{max}]$, etc.) in the spirit of the \citet{pawe15} study. Our goal was here to quantify, in a self-consistent way, the relative effects of the two potential dust-depletion mechanisms that are the surface energy constraint and the low-stirring dust-production imbalance. For the sake of clarity, and to clearly identify the mechanisms at play, we restricted our study to a reference wide disc and to the two illustrative reference cases of a sun-like and an A-type star. Test runs with a narrower, ring-like disc, have, however, been performed. They gave relatively similar results as to the relative imprint of the $s_{surf}$ and low-stirring effects on the PSD, although with a less pronounced waviness in the PSD's shape, which was to be expected, since small grains close to $s_{blow}$ will impact larger targets within a narrow ring at a lower velocity than they would have in an extended disc.

Despite this limited parameter exploration, we note however that our conclusions seem to agree with the numerical investigations of \citet{pawe15}, who were able to find a reasonable fit to the $s_{min}/s_{blow}$ trend observed on their 34-stars sample with the low-stirring-induced mechanism \emph{alone}. The main assumption for this fit to work is that, while the dynamical stirring of A stars should be of the order of $e\sim0.1$, it should decrease towards lower mass stars and be as low as $\sim0.01$ for solar-type objects. Such values might appear unrealistically low if we assume the classical view that debris discs are stirred by large lunar-to-Mars-sized large planetesimals \citep[e.g.,][]{theb09}. However, the level and the cause of stirring is still an open issue in present debris discs studies \citep[see discussion in][]{pawe15}. Moreover, we know of at least two discs, around the G0V-star HD207129 \citep{lohn12}, and the K2V-star HIP17429 \citep{schu14}, for which collisional modelling predicts <$e$> barely larger than 0.01.

\section{Summary}\label{summ}

Based on analytical considerations, the pioneering study of \citet{krij14} finds that the surface energy constraint, which limits the size of the smallest fragment produced after destructive collisions, could potentially affect the evolution of debris discs by limiting the smallest sizes of observable grains to a value $s_{surf}$ larger than the blow-out size $s_{blow}$.

Here we numerically quantify the importance of this mechanism by incorporating, for the first time, the surface energy constraint into a statistical code that follows the collisional evolution of a debris disc. Instead of relying on an analytical system-averaged value, we compute $s_{surf}$(i,j) for all impacting target-projectile pairs of sizes $s_i$ and $s_j$ in the disc. We consider two stellar types, sun-like and AV5, and two levels of stirring for the disc, <$e$>=0.075 and <$e$>=0.01. Our main results and conclusions are as follows:
\begin{itemize}
\item We confirm that, for all considered set-ups, there is a fraction of the ($s_i$,$s_j$) space for which $s_{surf}\geq s_{blow}$, especially for low <$e$> and/or low-mass stars.
\item However, the ($s_i$,$s_j$) regions of high $s_{surf}$ do \emph{not} coincide with the regions of high dust production. Indeed, most of the $\mu$m-sized dust is produced by impacts involving $\sim$3-20$s_{blow}$ targets and small $\lesssim$2-3$s_{blow}$ projectiles, for which $s_{surf}$ is, in general, smaller than the blow-out size. This is mainly due to the fact that impacts involving small grains are very energetic, because these particles are placed on high-$e$ orbits by radiation pressure 
\item Because of this discrepancy between high-$s_{surf}$ and dust-producing collisions, the global effect of the surface energy constraint is generally relatively limited. The only set-up for which it has a visible signature on the particle size distribution (PSD) is for a dynamically 'hot' disc around a solar-type star. But even there, the depletion of small dust does not exceed 30\%.
\item For the low <$e$>=0.01 cases, which should in principle be the most favourable to strong surface-energy constraints, the depletion of small grains never exceeds 12\% and is never able to change the sizes of particles that dominate the system's geometrical cross-section.
\item At such low-stirring levels, the system's PSD in the small size domain is, instead, totally dominated by another mechanism: the imbalance between small dust production and destruction rates identified by \citet{theb08} for low-$e$ discs. This imbalance creates a depletion of small grains that is at least one order of magnitude more pronounced than that caused by the surface-energy constraint, and it affects grains over a much wider size range.
\end{itemize}

Even if its effect is not as decisive as could be analytically expected, we do, however, recommend implementing the surface-energy constraint in collisional-evolution codes, as it might leave visible signatures in the low-$s$ end of PSDs for some star-disc configurations.

{}

\clearpage

\end{document}